\documentclass[aps,prl,twocolumn,groupedaddress,nofootinbib]{revtex4}
\usepackage[dvips]{graphicx}
\usepackage{color}
\usepackage{amsmath,amssymb,slashed}
\newif\ifdraft
\drafttrue
\newif\ifpreprint
\preprinttrue

\def\spa#1.#2{\left\langle#1\,#2\right\rangle}
\def\spb#1.#2{\left[#1\,#2\right]}

\newcommand{\eq}{\begin{equation}}
\newcommand{\eqe}{\end{equation}}
\newcommand{\eqa}{\begin{eqnarray}}
\newcommand{\eqae}{\end{eqnarray}}

\newcommand{\bea}{\begin{eqnarray}}
\newcommand{\eea}{\end{eqnarray}}
\newcommand{\dd}{\mathrm{d}}

\newbox\charbox
\newbox\slabox
\def\s#1{{      
        \setbox\charbox=\hbox{$#1$}
        \setbox\slabox=\hbox{$/$}
        \dimen\charbox=\ht\slabox
        \advance\dimen\charbox by -\dp\slabox
        \advance\dimen\charbox by -\ht\charbox
        \advance\dimen\charbox by \dp\charbox
        \divide\dimen\charbox by 2
        \raise-\dimen\charbox\hbox to \wd\charbox{\hss/\hss}
        \llap{$#1$}
}}

\begin{document}

\title{Universality in string interactions} 

\author{Yu-tin Huang$^a$,
Oliver Schlotterer$^{b}$
and Congkao Wen$^c$}
\affiliation{$^a$ Department of Physics and Astronomy, National Taiwan University, Taipei 10617, Taiwan, ROC}
\affiliation{$^b$ Max-Planck-Institut f\"ur Gravitationsphysik, Albert-Einstein-Institut,
14476 Potsdam, Germany,}
\affiliation{$^c$ I.N.F.N. Sezione di Roma ``Tor Vergata", Via della Ricerca Scientifica, 00133 Roma, Italy.}

\begin{abstract}
In this letter, we provide evidence for universality in the low-energy expansion of tree-level string interactions. More precisely, in the $\alpha'$-expansion of tree-level scattering amplitudes, we conjecture that the leading transcendental coefficient at each order in $\alpha'$ is universal for all perturbative string theories. We have checked this universality up to seven points and trace its origin to the ability to restructure the disk integrals of open bosonic string into those of the superstring. The accompanying kinematic functions have the same low-energy limit and do not introduce any transcendental numbers in their $\alpha'$-corrections. Universality in the closed-string sector then follows from the KLT-relations.
\end{abstract}

\maketitle

\section{Introduction}  

\vspace{-0.3cm}

One of the formidable challenges for a theory of quantum gravity is the construction of a gravitational S-matrix which respects unitarity at high energies. Perturbative string theories provide candidate solutions, as its four-point graviton S-matrix is exponentially suppressed in the high-energy limit for fixed-angle scattering~\cite{Gross, Amati:1987uf}. In fact, assuming tree-level causality \cite{Camanho:2014apa} and unitarity \cite{AHH} imposes stringent constraints, under which string theories provides the only known analytic solutions so far. 

Different string theories are understood to be equivalent through a web of strong-weak dualities which relate different orders in the perturbative expansion \cite{Witten:1995ex}. At tree level, however, the low-energy description in the form of an effective action with expansion in curvature tensors and covariant derivatives is largely unconstrained by string dualities. More precisely, the coefficients of these higher-dimensional operators are expected to be distinct for different string theories. Thus, if some of these coefficients turn out to be universal, it is then conceivable that such a phenomenon reflects a deeper principle in the theory of quantum gravity beyond the known dualities.

At low energies, closed-string theories yield an effective action that augments the Einstein-Hilbert term $S_{\rm EH}$ with higher-dimensional operators. At tree level, type-II superstring theories exhibit the following expansion in the inverse string tension (or cut-off scale) $\alpha'$,
\bea
S_{\rm eff}&=S_{\rm EH}-2\alpha'^3\zeta_3e^{-6\phi}R^4-\zeta_5\alpha'^5e^{-10\phi}D^4R^4\notag\\
& \ \ \ +\frac{2}{3}\alpha'^6\zeta_3^2e^{-12\phi}D^6R^4+\cdots \ , \label{effaction}
\eea
with Einstein-frame conventions for the dilaton couplings $e^{-n\phi}$. The ellipsis $\cdots$ represents loop-corrections and higher-order terms in $\alpha'$, while $D^nR^m$ schematically represent contractions of covariant derivatives and Riemann tensors. The tensor structure of each operator as well as its coefficient furnished by multiple zeta values (MZVs) 
\bea
\zeta_{n_1,n_2,\ldots,n_r} \equiv \! \! \sum_{0<k_1<k_2<\ldots<k_r}^{\infty} \! \! \frac{1}{k_1^{n_1} k_2^{n_2} \ldots k_r^{n_r}}
\label{MZV}
\eea
can be derived by expanding string-theory graviton amplitudes in $\alpha'$. MZVs can be conjecturally categorized according to their transcendental weight $n_1+n_2+\ldots+n_r$ and constitute a fruitful domain of common interest between high-energy physics and number theory. In fact, for type-II theories, the transcendental weight for each coefficient matches the order of $\alpha'$. This property will be referred to as {\em uniform transcendentality}, and it also exists for open strings in the type-I theory. The type-I effective action is now an expansion in non-abelian field-strength operators $\textrm{tr}(D^nF^m)$. In this light, uniform transcendentality for closed strings is inherited from open strings through the Kawai, Lewellen and Tye (KLT) relations~\cite{Kawai:1985xq}.

In this letter, we conjecture that {\it the leading transcendental coefficient at each order in the $\alpha'$-expansion of tree-level amplitudes is universal among all perturbative open- and closed-string theories.} We have explicitly verified this up to the seven-point level, and the conjectural all-multiplicity extension is discussed in a companion paper \cite{companion}. This remarkable property can be best understood by inspecting the world-sheet correlator of the open-string amplitudes. 

It was shown in~\cite{Mafra:2011nv} that the $n$-point tree amplitude of the open superstring can be cast into an $(n{-}3)!$ basis of disk integrals, each augmented by Yang-Mills tree amplitudes of different color-orderings. These basis integrals exhibit uniform transcendentality upon $\alpha'$-expansion, see e.g.~\cite{Broedel:2013aza} for a proof. We claim that bosonic open-string amplitudes can be cast upon the very same integral basis where -- in contrast to the superstring -- the accompanying functions of the kinematic data depend on $\alpha'$. Apart from the Yang-Mills trees recovered in their low-energy limit $\alpha'\rightarrow 0$, the $\alpha'$-corrections of the kinematic functions exclusively involve rational numbers upon Taylor-expansion, i.e.~they do not carry any transcendental weight. Hence, the resulting $\alpha'$-expansion of the bosonic string amplitude will have the same leading transcendental pieces as found for the superstring.

The same property can be extended to closed strings by utilizing the KLT-relations~\cite{Kawai:1985xq}, which assemble closed-string tree amplitudes from products of two open-string trees. The accompanying $\sin$-functions with $\alpha'$-dependent arguments do not alter the uniform transcendentality of the type-II theory. 
Different double-copies of~open bosonic strings and superstrings give rise to three different closed-string theories -- bosonic, heterotic and type-II superstrings. Their tree amplitudes are governed by a universal basis of  $(n{-}3)!\times (n{-}3)!$ integrals of uniform transcendentality inherited from the open-string constituents. Only the kinematic coefficients differ between the theories, where the additional $\alpha'$-corrections specific to open bosonic strings do not introduce any transcendental weight and thereby do not affect the leading-transcendental piece. This completes the argument for universality in closed-string interactions, namely for the $\mathcal{O}(\alpha'^n)$ order of the effective action, the weight-$n$ coefficient is universal for all perturbative closed-string theories.

\vspace{-0.4cm}

\section{Open-string amplitudes}

\vspace{-0.3cm}

{\bf A. The open superstring:} 
The tree-level amplitude for $n$ gluon-multiplet states in open superstring theory can be conveniently written as~\cite{Mafra:2011nv}
\bea \label{eq:generalopen}
{\cal A}^S(1,2_\rho, \ldots, (n{-}2)_{\rho }, n{-}1,n;\alpha') = 
\sum_{\sigma \in S_{n-3}} F_\rho{}^{\sigma}(\alpha')\nonumber\\
\times A_{\rm YM} (1, 2_{\sigma }, \ldots, (n{-}2)_{\sigma }, n{-}1, n) \ ,
\eea
where ${\cal A}^S$ and $A_{\rm YM}$ indicate color-ordered amplitudes of the superstring and super Yang-Mills field theory, respectively. Moreover, $\rho,\sigma$ with $j_\rho \equiv \rho(j)$ denote the $(n{-}3)!$ distinct permutations with legs $1,n{-}1, n$ held fixed, and $F_{\rho}{}^{\sigma}(\alpha')$ are disk integral that capture the $\alpha'$-dependence,
\bea
F_\rho{}^{\sigma}(\alpha') \equiv 
\! \! \! \! \! \! \! \! \! \! \! \! \! \! \! \! \! \! \! \! \! \! \! \! \! \! \int \limits_{0\leq  z_{2_\rho} \leq z_{3_\rho}\leq \ldots \leq z_{(n-2)_\rho} \leq 1 }\! \! \! \! \! \! \! \! \! \! \! \! \! \! \! \! \! \! \! \! \! \! \! \! \! \!
  \dd z_2 \ldots \dd z_{n-2} \, \prod^n_{i<l} |z_{il}|^{s_{il}} \sigma \Big\{ 
 \prod^{n-2}_{k=2} \sum^{k-1}_{m=1} {s_{mk}  \over z_{km}} \Big\} \, , 
 \label{oli0}
\eea
with $z_{ij} \equiv z_i - z_j$. We fix the SL$(2)$ symmetry of the disk by setting $(z_1,z_{n-1},z_n)=(0,1,\infty)$, and we use dimensionless Mandelstam invariants
\bea
 s_{ij \ldots l} \equiv \alpha'(k_i+k_j+ \ldots +k_{l})^2 \ .
\eea
When viewed as an $(n{-}3)! \! \times \!(n{-}3)!$ matrix, the row- and column indices $\rho$ and $\sigma$ of $F_\rho{}^{\sigma}$ label different integration domains and integrands, respectively, where $\sigma$ acts on the subscripts within the curly bracket in (\ref{oli0}). Note that the field-theory limit is recovered as $F_\rho{}^{\sigma}(\alpha') = \delta_\rho{}^{\sigma} +{\cal O}(\alpha'^2)$, and the $(n{-}3)!$-vector in (\ref{eq:generalopen}) furnishes a basis of string subamplitudes under monodromy relations \cite{BjerrumBohr:2009rd, Stieberger:2009hq}.

The $\alpha'$-expansion of the integrals in (\ref{oli0}) yields MZVs (\ref{MZV}) whose transcendental weight matches the degree of the accompanying polynomials in $s_{ij}$. Since $A_{\rm YM}$ do not depend on $\alpha'$, uniform transcendentality of the integrals propagates to the disk amplitude (\ref{eq:generalopen}). Initially addressed via hypergeometric functions \cite{Oprisa:2005wu}, the $\alpha'$-corrections of $F_\rho{}^{\sigma}(\alpha') $ at any multiplicity can be recursively generated from the Drinfeld associator \cite{Broedel:2013aza}.


Once undoing the above choice of SL(2) frame, the functions (\ref{oli0}) can be identified as a superposition of $(n{-}3)!$ ``single-cycle'' disk integrals,
\bea
\label{oli01}
Z_\rho(1_\sigma,2_\sigma,{\ldots},n_\sigma) \equiv
\int \frac{  \dd \mu_n(\rho) }{ \sigma(z_{12}z_{23}  \ldots z_{n1})}
 \ ,
\eea
where $\sigma$ and $\rho$ now act on all external legs in the integrand and the integration domain, respectively, and the measure is given by
\bea
 \int \dd \mu_n(\rho) \equiv \! \! \!  \! \! \!  \! \! \!  \! \! \! 
 \int \limits_{  -\infty <  z_{1_\rho} \leq z_{2_\rho}\leq \ldots \leq z_{n_\rho  } < \infty}
  \! \! \!  \! \! \!  \! \! \!  \! \! \! 
 \frac{ \dd z_1 \, \dd z_2 \ldots \dd z_{n} }{\textrm{vol} (\textrm{SL}(2))}  \,  \prod^n_{i<l} |z_{il}|^{s_{il}}\ .
\label{meas}
\eea
The integral reductions performed in \cite{Mafra:2011nv} rely on partial-fraction manipulations and integrations by parts (IBP) among $Z_\rho(1_\sigma,{\ldots},n_\sigma) $. At fixed $\rho$, these integral relations for different choices of $\sigma$ can be identified with the KK- and BCJ-relations \cite{Bern:2008qj} of $A_{\rm YM}(\ldots)$ \cite{Broedel:2013tta}. However, as already exploited in a superstring context \cite{Mafra:2010gj, Mafra:2011nv}, IBP additionally allows to address closed subcycles of $z_{ij}$ in the integrand such as double poles $z_{ij}^{-2}$. Extending these techniques to gluon amplitudes of the bosonic string yields our main result to be reported in the following.

\medskip
{\bf B. The bosonic open string:} 
The tree-amplitude prescription for $n$-gluon scattering in the bosonic string introduces significantly more rational functions of $z_{ij}$ of suitable SL$(2)$ weight than captured by the single cycles in (\ref{oli01}). Still, repeated use of IBP is expected to reduce all of them to the single-cycle form and thereby to the {\it same integral basis} as seen in (\ref{eq:generalopen}) and (\ref{oli0}), e.g.
\begin{eqnarray} \label{eq:4pt}
\int   \frac{ \dd \mu_{4} (\rho)  }{ z_{14}^2 z_{23}^2}
  = \frac{ s_{12}  Z_\rho(1,2,3,4) }{1-s_{23}} \,.
\end{eqnarray}
The denominator on the right-hand side signals tachyon exchange specific to the bosonic string and can be expanded as a geometric series $(1-s_{ij})^{-1}= \sum_{k=0}^{\infty}s_{ij}^k$. In a superstring context, the OPE among supersymmetric vertex operators guarantees that tachyon poles as in (\ref{eq:4pt}) are suppressed by numerators $1-s_{ij}$, see e.g.~\cite{Mafra:2010gj, Mafra:2011nv}. Extending the integral reduction along the lines of (\ref{eq:4pt}) to arbitrary multiplicity leads us to conjecture the following structure for the $n$-gluon tree in bosonic string theory:
\bea \label{oli04}
{\cal A}^B(1,2_\rho, \ldots, (n{-}2)_{\rho }, n{-}1,n;\alpha') = 
\sum_{\sigma \in S_{n-3}} F_\rho{}^{\sigma}(\alpha')\nonumber\\
\times B(1, 2_{\sigma }, \ldots, (n{-}2)_{\sigma }, n{-}1, n;\alpha') \ .
\eea
In comparison to the superstring result (\ref{eq:generalopen}), the kinematic factors $A_{\rm YM}(\ldots)$ are replaced by more general and $\alpha'$-dependent objects $B(\ldots;\alpha')$. Both of them are rational functions of $s_{ij}$ and multilinear in the polarizations $e_j$ entering via $(e_i\cdot e_j)$ and $(e_i \cdot k_j)$, and crucially do not carry any transcendental weights. Upon $\alpha'$-expansion, the leading term reproduces Yang-Mills tree amplitudes, and is therefore identical to that of the superstring, i.e.
\bea \label{oli91}
B(1,\!\ldots \!,n;\alpha') = A_{\rm YM}(1,\!\ldots \!,n)  + \sum_{k{=}1}^{\infty}(2\alpha')^k B_k(1,\!\ldots \!,n) \, .
\eea
At generic multiplicity $n$, the $B_{k}(\ldots)$'s have homogeneity degree $4{-}n{+}2k$ in momenta. The simplest instances of the subleading terms occur at the three-point level and signal the $F^3$ interaction specific to the bosonic string,
\bea \label{oli05}
B_1(1,\!2,\!3)= (e_1 \! \cdot \! k_2)(e_2 \! \cdot \! k_3)(e_3 \! \cdot \! k_1)  \, , \ \  B_{k\geq 2}(1,\!2,\!3) = 0
 \, .
\eea
The higher-point case requires integral reductions as in (\ref{eq:4pt}), and the resulting geometric series yield non-zero $B_{k}(\ldots)$ for any value of $k$. In the case of $n{=}4$, we find
\bea
&\! \! \! \! \! \! \! \! \! \! B(1,2,3,4;\alpha')= A_{\rm YM}(1,2,3,4) + (2 \alpha')^2  \label{Bk4} \\ 
&\ \ \ \  \times s_{13} \left[ \left( 
{ f_{12} f_{34} \over s_{12}^2 (1 - s_{12}) } + \textrm{cyc}(2,3,4) \right) - 
{ g_1 g_2 g_3 g_4 \over s_{12}^2 s_{13}^2 s_{14}^2 } \right] \, ,
\notag
\eea
with gauge invariant constituents $f_{ij} \equiv (e_i \cdot  e_j)  (k_i \cdot k_j)- (k_i \cdot e_j) ( k_j \cdot e_i)$ and $ g_i \equiv (k_{i{-}1} \cdot e_i)  s_{i , i{+}1} {-} (k_{i{+}1} \cdot e_i) s_{i{-}1, i}$. Note that both $s_{ij}$ and $g_i$ carry a power of $\alpha'$ when extracting the $B_k(1,2,3,4)$'s from the second line of (\ref{Bk4}).

It is crucial to note that no MZVs or transcendental weight accompany the $\alpha'$-dependence from $B(\ldots;\alpha')$. Given the uniform transcendentality of the $F_{\rho}{}^{\sigma}(\alpha')$ and the absence of negative powers of $\alpha'$ in the kinematic factor (\ref{oli91}), the transcendental weight cannot exceed the accompanying order in $\alpha'$ within the bosonic-string amplitude. At fixed order in $\alpha'$, the leading-transcendental part of the open bosonic string follows from picking up $B(\ldots;\alpha')\rightarrow A_{\rm YM}(\ldots)$ in (\ref{oli91}) and therefore agrees with the superstring amplitude. \textit{This leads to the conclusion that the leading-transcendental pieces of the tree-level $\alpha'$-expansion and the resulting $\textrm{tr}(D^m F^n)$ interactions are universal in open-string theories.}

 \medskip
{\bf C. BCJ-symmetries of the kinematic factors:} Although the kinematic factors $B_k(\ldots)$ in (\ref{oli91}) differ from $A_{\rm YM}(\ldots)$ in tensor structure and mass dimension, we will now argue that they obey the same KK- and BCJ-relations \cite{Bern:2008qj}. The universal monodromy relations \cite{BjerrumBohr:2009rd, Stieberger:2009hq} among bosonic-string subamplitudes have to hold separately at each order in $\alpha'$ and along with each transcendentality. Hence, inserting (\ref{oli04}) into the lowest-transcendentality pieces of the monodromy relations and identifying $B_0(\ldots) \equiv A_{\rm YM}(\ldots)$ yields
\begin{align}
0&=B_k(1,2,\ldots,n)+  B_k(2,1,3, \ldots,n)  +  B_k(2,3,1, \ldots,n)     \notag \\
&\ \ +   \ldots +B_k(2,3,\ldots,n-1,1,n) 
\label{eq:BCJ} \\
0&= s_{12}B_k(2,1,3, \ldots,n)   +  (s_{12}\! +\! s_{13})B_k(2,3,1,4,\ldots,n) \notag \\
& \ \ +  \ldots + (s_{12}\!+\!s_{13}\!+\!\ldots\!+\!s_{1,n-1}) B_k(2,3,\ldots,n-1,1,n) \notag
\end{align}
for any value of $k$. The idea of imposing monodromy relations order by order has been exploited in \cite{Broedel:2012rc} to derive BCJ-relations for subamplitudes of the $F^3$ operators as well as the supersymmetrized $D^2 F^4+F^5$. Moreover, a general argument for the entire gauge sector of the heterotic string has been given in \cite{Stieberger:2014hba}. By the same reasoning, (\ref{eq:BCJ}) can be extended to an infinity of $\alpha'$-corrections
\begin{align}
&B_k^{j_1j_2\ldots j_p}(1, 2_{\sigma }, \ldots, (n{-}2)_{\sigma }, n{-}1, n) \equiv  \sum_{\tau \in S_{n-3}}  \label{deformation} \\
& \ \ \ (M_{j_1} M_{j_2} \ldots M_{j_p})_{\sigma}{}^{\tau} 
B_k(1, 2_{\tau }, \ldots, (n{-}2)_{\tau }, n{-}1, n) \ ,
\notag
\end{align}
labelled by $j_i \in 2\mathbb N+1$. The $(n{-}3)! \times (n{-}3)!$ matrix $M_{j}$ is the coefficient of $\zeta_j$ when casting the $\alpha'$-expansion of $F_\rho{}^\sigma$ in (\ref{oli0}) into a conjectural basis of MZVs w.r.t.~rational numbers $\mathbb Q$ \cite{Schlotterer:2012ny}. The entries of $M_j$ are degree-$j$ polynomials in $s_{pq}$, see \cite{WWW} for examples at multiplicity $n \leq 7$. Note that the symmetry properties (\ref{eq:BCJ}) of $B_k(\ldots)$ and their deformations $B^{j_1\ldots j_p}_k(\ldots)$ in (\ref{deformation}) are inevitable to verify permutation invariance of the world-sheet integrand for the bosonic-string amplitude along with each transcendentality and order in $\alpha'$. 

\medskip
{\bf D. Supporting evidence:} To confirm the central conjecture (\ref{oli04}) implying our universality results, one must prove that the complete bosonic-string integrand including multi-cycle generalizations of (\ref{oli01}) can be reduced to the single-cycle case. While a systematic all-multiplicity analysis is relegated to future work \cite{companion}, the following IBP identities provide substantial support.

At five points, after partial-fractional manipulations, we need following two identities in addition to reduce all the integrals to a single-cycle basis (\ref{oli01}):
\begin{align} 
\int  \frac{ \dd \mu_5(\rho)}{z_{23}^2\,(z_{15}z_{54}z_{41})} &= \frac{s_{12} Z_\rho(1,2,3,5,4)-(1\leftrightarrow 4) }{s_{23}-1}\,,  \notag \\
\int \frac{\dd \mu_5(\rho) \,  z_{25} }{z_{23}^2 z_{15}^2 z_{24} z_{45} } &= 
 \frac{ s_{13} Z_\rho(1,3,2,4,5) }{ 1- s_{51} }      \label{eq:5pt} \\
 + \frac{ s_{14} }{1-s_{51}}&
\left[
\frac{s_{12} Z_\rho(1,2,3,5,4)-(1\leftrightarrow 4) }{s_{23}-1}
\right]\,.
\notag 
\end{align}
The resulting form of $B(1,2,3,4,5;\alpha')$ is rather lengthy, and an auxiliary {\it mathematica} notebook containing the full expression is attached to the arXiv submission.

To arrive at (\ref{oli04}) at six points, we find that after partial-fraction manipulations, besides the single-cycle basis we encounter integrands of following forms, 
 \begin{align}
&\frac{ 1 }{(z_{23}z_{34}z_{42}) \, (z_{15}z_{56}z_{61})} \,, \quad \frac{1 }{ z_{23}^2 \,(z_{14} z_{46} z_{65} z_{51})} \,,  \\
&{ 1 \over z_{12}^2 z_{34}^2 z^2_{56} } \, , \quad { z_{36} \over z_{23}^2 z_{56}^2 z_{13} z_{14} z_{46} } \, , \quad 
{ z_{26} \over z_{23}z_{34}z_{42} z_{56}^2 z_{12} z_{16} } \,. \nonumber
\end{align}
We have checked that indeed all the above six-point integrals can be reduced to single-cycle integrals via IBP,~e.g.
\begin{align}
\int \frac{(s_{234}-1)\, \dd \mu_6(\rho) }{(z_{23}z_{34}z_{42}) \, (z_{15}z_{56}z_{61})}
&= s_{13} Z_\rho(1,3,4,2,6,5) \label{eq:6pt} \\ 
 - s_{35} &Z_\rho(1,6,2,4,3,5) - (3\leftrightarrow 4)\, , \notag
\end{align}
and the analogous seven-point checks to arrive at (\ref{oli04}) have been performed as well.
Note that all identities of (\ref{eq:4pt}), (\ref{eq:5pt}) and (\ref{eq:6pt}) can alternatively be derived by imposing linearized gauge invariance under $e_j\rightarrow k_j$, and the same is believed to hold for the integral reduction at arbitrary multiplicity.

\vspace{-0.4cm}

\section{Closed-string amplitudes}

\vspace{-0.3cm}

Closed-string amplitudes at tree level can be obtained from squares of open-string amplitudes through the KLT-relations \cite{Kawai:1985xq}. The accompanying sin-functions of $\pi s_{ij}$ conspire with the $\alpha'$-expansion of the open string such as various MZVs including all $\zeta_{2n}$ cancel in a suitable basis w.r.t.~$\mathbb Q$ \cite{Stieberger:2009rr, Schlotterer:2012ny}. These selection rules were identified in \cite{Stieberger:2013wea} with the single-valued projection of MZVs~\cite{Schnetz:2013hqa}.

\medskip
{\bf A. The closed superstring:}  A representation of the massless closed-superstring tree ${\cal M}^S_n$ which manifests the effect of these cancellations has been firstly given in~\cite{Schlotterer:2012ny}:
\begin{align}
{\cal M}^S_n (\alpha') &= \!\!\! \sum_{\sigma,\rho,\tau \in S_{n-3}} \! \! \! \! \!  \! \! \tilde A_{\rm YM}(1,2_\sigma,\ldots,(n{-}2)_\sigma,n,n{-}1)
(S_0)_\sigma{}^\rho \label{oli08}
\notag\\ 
& \! \! \! \!\! \! \! \! \times  \ G_\rho{}^\tau(\alpha') A_{\rm YM}(1,2_\tau,\ldots,(n{-}2)_\tau,n{-}1,n) \ .
\end{align}
The polarizations of the type-II supergravity multiplets stem from tensor products of the gauge-multiplet polarizations in $\tilde A_{\rm YM}$ and $A_{\rm YM}$. The matrix $S_0$ has entries of order $(k_i{\cdot} k_j)^{n-3}$ and appears in the momentum-kernel representation \cite{BjerrumBohr:2010hn} of the KLT-formula for supergravity trees \cite{Bern:1998sv}. The matrix $G_\rho{}^\sigma(\alpha')$ in (\ref{oli08}) carries the entire $\alpha'$-dependence and takes the form \cite{Schlotterer:2012ny}
\bea
G(\alpha') = 1 + 2 \zeta_3 M_3 + 2 \zeta_5 M_5 + 2 \zeta_3^2 M_3^2 + {\cal O}(\alpha'^7) \ .
\label{clexpa}
\eea
The matrices $M_3$ and $M_5$ have been introduced with (\ref{deformation}), and $(n\leq 7)$-point examples are available from \cite{WWW}. Together with the polarization-dependence from the super Yang-Mills trees, they encode the tensor contractions of the $D^n R^m$ operators in the tree-level effective action to the orders displayed in (\ref{effaction}).

Given the ubiquitous matrix products with summations over permutations in $S_{n-3}$, we will drop indices henceforth and rewrite (\ref{oli08}) in the condensed notation
\bea
{\cal M}^S_n (\alpha') &= \tilde A_{\rm YM}\cdot S_0\cdot G(\alpha')\cdot  A_{\rm YM} \ ,
\label{oli09a}
\eea
where the vectors $\tilde A_{\rm YM}$ and $A_{\rm YM}$ are understood to be in the different $(n{-}3)!$-bases spelt out in (\ref{oli08}).

\medskip
{\bf B. Universality for closed string theories:} As exploited in \cite{Stieberger:2014hba} for the heterotic string, the above structure and $\alpha'$-expansion of type-II closed-string amplitudes are a property of the world-sheet integrals when two copies of the integrands in (\ref{oli0}) are integrated over the sphere. Accordingly, the results on the integrals can be imported in further contexts such as gravitational tree amplitudes ${\cal M}_n^{H}$ or ${\cal M}_n^{B}$ in the heterotic or the closed bosonic string which rest on one or two copies of the bosonic-string integrand in (\ref{oli04}). The only modification as compared to the superstring (\ref{oli08}) is an exchange of $A_{\rm YM}(\ldots)\!\leftrightarrow \!B(\ldots,\alpha')$,
\begin{align}
{\cal M}_n^{H} (\alpha') &= \tilde A_{\rm YM}\cdot S_0\cdot G(\alpha')\cdot  B(\alpha') 
\label{oli13} \\
{\cal M}_n^{B} (\alpha') &= \tilde B(\alpha')\cdot S_0\cdot G(\alpha')\cdot  B(\alpha')  \ ,
\label{oli14}
\end{align}
where the same bases of color-orderings spelt out in (\ref{oli08}) are used for $\tilde B(\alpha'),\tilde A_{\rm YM}$ and $B(\alpha'),A_{\rm YM}$, respectively. Clearly, the Einstein-Hilbert interaction can be recovered at leading order in $\alpha'$ where $G(\alpha')\rightarrow 1$, $B(\alpha')\rightarrow A_{\rm YM}$ and $\tilde B(\alpha')\rightarrow \tilde A_{\rm YM}$, and the  $R^2$-correction at subleading order in $\alpha'$ is recovered from instead setting $B(\alpha')\rightarrow 2\alpha' B_1$ in (\ref{oli13}). Note that, by level-matching,~the tachyonic poles of the form $(1-s_{ij\ldots l})^{-1}$ in $B(\alpha')$ of the heterotic amplitude (\ref{oli13}) are cancelled by corresponding zeros in the entries of $G(\alpha')$ as $s_{ij \ldots l} \rightarrow 1$. The structure of (\ref{oli13}) is expected to capture multitrace interactions in the gauge sector of the heterotic string under appropriate replacement of $B(\alpha')$, see \cite{companion} for further details.

In complete analogy to the open bosonic string, it is natural to organize (\ref{oli13}) and (\ref{oli14}) in a double-expansion w.r.t.~$\alpha'$ and transcendental weight. While the kinematic factors $B(\ldots;\alpha')$ with an expansion as in (\ref{oli91}) only involve rational coefficients, the $\alpha'$-corrections from $G(\alpha')$ still enjoy uniform transcendentality. At fixed order in $\alpha'$, the leading-transcendentality part is again obtained by truncating $B(\ldots,\alpha')  \rightarrow A_{\rm YM}(\ldots)$ and therefore identical in (\ref{oli09a}), (\ref{oli13}) and (\ref{oli14}). {\it Hence, we have shown that, at leading transcendentality, gravitational tree-level interactions are universal to the bosonic, heterotic and type-II closed-string theories.}

\vspace{-0.4cm}

\section{Conclusions}

\vspace{-0.3cm}

In this letter, tree-level amplitudes in all perturbative open- and closed-string theories are argued to have universal leading-transcendental parts in their $\alpha'$-expansions. Manifest universality can be achieved by casting the world-sheet correlators of the bosonic open string into the same basis of disk integrals as the superstring, augmented with $\alpha'$-dependent kinematic factors. We have explicitly shown that such a reorganization can be achieved up to seven points, and the conjectural all-multiplicity extension is relegated to future work \cite{companion}. Generalizations to closed-string interactions in bosonic, heterotic and type-II theories directly follow from the KLT-relations. These universality results have greatly 
facilitated the construction of matrix elements for counterterms in half-maximal supergravity~\cite{Huang:2015sla}. 

It would be interesting to apply the same organizing principles to massive-state scattering. We expect the same basis of disk integrals to capture tree amplitudes among any combination of massive open-string resonances. Moreover, the structure of (\ref{oli09a}) is believed to apply to closed-string trees among massive resonances upon appropriate replacements of $A_{\rm YM}$ and $\tilde A_{\rm YM}$.

\vspace{-0.4cm}

\section{Acknowledgements}

\vspace{-0.3cm}

We are grateful to Johannes Br\"odel, Paolo Di Vecchia, Michael Green and Henrik Johansson for inspiring discussions and valuable comments on a draft of the manuscript. Also, Massimo Bianchi and Andrea Guerrieri are thanked for enlightening discussions. Y-t.~H.~is supported by MOST under the grant No.~103-2112-M-002-025-MY3, and O.~S.~is grateful to the Universit\`a di Roma Tor Vergata for kind hospitality during finalization of this work. 

\end{document}